\documentclass[12pt,letterpaper]{article}
\usepackage{iftex}
\ifPDFTeX
  \usepackage[T1]{fontenc}
  \usepackage[utf8]{inputenc}
\fi
\usepackage{mathptmx}
\usepackage[margin=1in]{geometry}
\usepackage{setspace}
\usepackage{amsmath,amssymb}
\usepackage{graphicx}
\usepackage{booktabs}
\usepackage{array}
\usepackage{caption}
\usepackage[authoryear]{natbib}
\setcitestyle{aysep={}}
\usepackage{float}
\usepackage{algorithm}
\usepackage{algpseudocode}
\usepackage[hidelinks]{hyperref}
\graphicspath{{figures/}}
\newcommand{\se}{\operatorname{se}}
\newcommand{\sd}{\operatorname{sd}}
\newcommand{\sgn}{\operatorname{sign}}
\newcommand{\pdir}{p_{\mathrm{dir}}}

\newif\ifidentified
\identifiedfalse   

\begin{document}
\doublespacing
\begin{center}
{\Large\bfseries Pseudo-Incrementality Testing: Measuring Advertising Lift
from Naturally Occurring Interventions\par}
\vspace{0.5em}
{Georgios Filippou\par}
{\small Trinity College Dublin,  School of Computer Science and Statistics, Dublin, Ireland\par}
{\small Correspondence: filippog@tcd.ie\par}
\vspace{0.4em}
{Boi Mai Quach\par}
{\small Trinity College Dublin,  School of Computer Science and Statistics, Dublin, Ireland\par}
\vspace{0.4em}
{Ashish Kumar Jha\par}
{\small Trinity College Dublin, Trinity Business School, Dublin, Ireland\par}
\vspace{0.4em}
\end{center}

\section*{Abstract}
We develop a method for measuring the incremental effect of advertising when randomized experiments are unavailable. Firms generate abrupt interventions in their own marketing as a byproduct of operations: budgets are cut, channels launch, programs pause. We propose a two-stage procedure that treats these events as quasi-experiments. The first stage discovers and dates interventions by exact Bayesian run-length inference on marketing activity series. The second stage runs a causal impact analysis against a variance-constrained structural time series counterfactual, with simulation-based inference, five qualification conditions, and a closed-form power bound. We validate the procedure on the two experimental benchmarks that Meta released with its GeoLift framework. The method identifies the day of intervention exactly in both cases. On the experiment where advertising was removed, it recovers the effect within 6.2 percent of the experimental estimate (implied return of 1.50 versus 1.60). On the experiment where advertising was added, its 90 percent interval covers the experimental estimate and excludes zero. Of six estimators evaluated on the same released data, ours is the only one that requires no geographic panel and produces intervals that are both correct on both experiments and narrow enough to act on.

\noindent\textit{Keywords:} causal inference; change point detection;
counterfactual prediction; quasi-experiments; structural
time series

\section{Introduction}\label{sec:intro}
Advertisers measure the causal effect of marketing spending with randomized
experiments where they can. Platform conversion lift studies randomize users;
geographic experiments randomize markets \citep{vaver11,johnson17}. The
experimental gold standard is well earned: observational estimates of
advertising effects diverge from experimental benchmarks frequently,
unpredictably, and in both directions, so that no fixed correction rescues
them \citep{gordon19,blake15}. The advice advertisers receive, from the
platforms that build lift infrastructure and from the literature that audits
it, is accordingly uniform: experiment where possible. We agree with the
advice. This article is about the large and permanent territory where it
cannot be followed.

Let $y_t$ denote a business outcome and $x_{k,t}$ the activity of marketing
channel $k$. Suppose channel $k$ undergoes an abrupt, sustained change at a
date $t^{*}$ that appears in no record the analyst can consult. We call such
an event a naturally occurring intervention. Writing $y_t(0)$ for the outcome
path that would have prevailed absent the intervention, the object of
interest is the cumulative lift over a post-window of $W$ periods,
\begin{equation}\label{eq:estimand}
\tau \;=\; \sum_{t \in \mathrm{post}} \bigl[\, y_t - y_t(0) \,\bigr],
\qquad \mathrm{post} = \{\, t^{*}, \ldots, t^{*} + W - 1 \,\},
\end{equation}
together with the return per unit of spending change, $\rho = \tau / \Delta
s$. Two features distinguish this problem from standard program evaluation.
First, the intervention date is not known to the analyst; interventions
accumulate in operational data faster than institutional memory records them,
and the estimation problem includes finding them. Second, no randomization
and typically no geographic panel are available; the only information for
constructing $y_t(0)$ is the set of series co-observed with the outcome,
namely the remaining marketing channels, calendar structure, and demand
covariates.

Why estimate $\tau$ from found interventions rather than designed ones?
Because the supply of designed ones is rationed by arithmetic. Consider an
experiment in which spending $S$ is applied to a channel over $W$ periods and
the true return per unit of spending is $r$, so that incremental revenue $rS$
accrues against outcome noise with per-period standard deviation $\sigma$. A
comparison of treated and control totals estimates $rS$ with standard error
$\sigma\sqrt{2W}$, so the design carries a $t$-statistic of approximately
\begin{equation}\label{eq:power}
t \;\approx\; \frac{r S}{\sigma \sqrt{2W}},
\end{equation}
and inverting the formula at conventional size and power gives the minimum
detectable return, proportional to $\sigma\sqrt{2W}/S$. The quantity that
decides everything is spending measured against accumulated outcome noise,
and for most advertisers most of the time it is small. \citet{lewisrao15}
develop this argument and document its force: in a set of large display
advertising field experiments, samples in the millions of individuals
routinely failed to determine whether the return to advertising was even
positive. \citet{shapiro21} reach the complementary conclusion from the
observational side, estimating television advertising elasticities across 288
brands and finding a distribution concentrated near zero, effects of a size
that conventionally powered designs cannot resolve. The consequence in
practice is not an absence of experimentation but a waste of it: firms run
underpowered tests, obtain insignificant results at full media cost, and draw
either no conclusion or the wrong one.

The premise of this article is that the historical data most firms already
hold contain interventions of exactly the kind an experimentalist would
design, at spending changes larger than any test budget would authorize, and
at zero incremental cost, because they have already happened. Budgets are cut
in downturns and restored afterward; channels are launched, paused during
agency transitions, and restarted; platforms are abandoned and readopted.
What these events lack is documentation and a control group. We propose a
two-stage procedure, pseudo-incrementality testing, that supplies both. Stage
one applies exact Bayesian online change point detection \citep{adams07} to
every channel's activity series, maintaining the full posterior distribution
of the time since the last structural break; a break is declared and dated
when the accumulated posterior mass on short run lengths crosses a threshold.
Stage two runs a causal impact analysis of the dated intervention on the
outcome variable: a structural time series counterfactual
\citep{harveytodd83,brodersen15} in which the level and slope disturbances
are pinned to small fractions of pre-window outcome variation, so that
identification runs through co-observed regressors rather than through a
drifting intercept, with simulation-based inference. The output is graded:
five qualification conditions decide whether the estimate may be treated as
evidence, and a closed-form power bound identifies channels that cannot be
tested at a given aggregation at all.

To the best of our knowledge, no existing procedure combines these elements.
Intervention analysis assumes the intervention date is known
\citep{boxtiao75}. Causal impact estimation assumes a known date and an
unconstrained local level, which we show produces intervals too wide to be
decision-relevant in this application \citep{brodersen15}. Synthetic control
and difference-in-differences require a panel of untreated units, typically
geographic, which most advertisers do not possess \citep{abadie21,kaul22}.
Geo-experiment analysis requires the randomized panel it was designed around
\citep{vaver11}. Change point detection locates breaks but attaches no causal
estimand to them \citep{page54,basseville93,perron21}. Our contributions are
accordingly: (a) a formal two-stage estimator linking run-length detection to
counterfactual lift estimation; (b) an identification analysis with named
assumptions and a characterization of a donor-contamination bias, with the
resulting design rule; (c) a qualification and power framework that converts
the estimator into an evidence-grading procedure; (d) an external validation
against the two experimental benchmark analyses published by Meta, including
an inverse experiment, with a six-estimator comparison on the same released
data; and (e) simulation experiments with exact ground truth mapping the
operating characteristics.

The validation previews as follows, and Section~\ref{sec:verdict} states the
verdict plainly. On Meta's published GeoLift benchmark, with every tuning
constant fixed in advance, stage one dates the intervention to September 1
exactly, from the spending series alone. On the inverse experiment, in which
advertising was removed from ten markets, the estimated lift is within
6.2 percent of the experimental estimate, and on the economic scale that
managers use the estimator prices the channel at a return of 1.50 against the
experiment's 1.60. On the positive experiment the 90 percent interval covers
the experimental estimate and excludes zero, with a point estimate 19.9
percent conservative. Across six estimators on the same released data, it is
the only one that is feasible without a geographic panel and produces
intervals that are correct on both experiments and narrow enough to act on.
We are explicit about what these percentages mean and about where competitors
do better, and equally explicit that a randomized experiment analyzed with
its own machinery remains the superior instrument wherever it exists.

Section~\ref{sec:setup} presents the setup and assumptions.
Section~\ref{sec:detect} develops the detection mathematics.
Section~\ref{sec:cf} develops the counterfactual model.
Section~\ref{sec:impact} assembles the causal impact analysis and its
qualification. Section~\ref{sec:relation} relates the procedure to existing
methods. Section~\ref{sec:bench} describes the benchmark data in detail and
reports the validation. Section~\ref{sec:sim} maps operating characteristics
in simulation. Section~\ref{sec:verdict} states the verdict and
Section~\ref{sec:conc} concludes. Replication code and all datasets are
provided as supplementary material.

\section{Setup and Assumptions}\label{sec:setup}
Data consist of $T$ periods of an outcome $y_t > 0$, channel activities
$x_{k,t} \ge 0$ for $k = 1, \ldots, K$, spending $s_{k,t}$, and covariates
$d_t$ external to marketing. Periods are days or weeks. For a focal channel
$k$, define the co-observed information set $z_t = (\, \{ \log(1 + x_{j,t})
\} \text{ for } j \ne k$, calendar terms including day-of-week structure on
daily data, the outcome lagged one year where the sample covers it, $d_t
\,)$. Potential outcomes $y_t(0)$ are defined with respect to the focal
intervention only. The estimand is \eqref{eq:estimand}; the return is $\rho =
\tau / \Delta s$ with $\Delta s = \sum_{\mathrm{post}} ( s_{k,t} - \bar
s_{k,\mathrm{pre}} )$. One requirement on $d_t$ deserves emphasis: it should
contain at least one demand measure the intervention cannot touch. Footfall,
arrivals, and weather qualify; so does the outcome aggregated over regions
whose spending never changed, which is available whenever the outcome is
recorded by region, because the unchanged regions are identifiable from the
spending data alone. This last construction is exactly the one the benchmark
validation of Section~\ref{sec:bench} uses.

\medskip\noindent\textbf{Assumption A} (Piecewise Stationary Activity).
\textit{Each activity series $x_{k,t}$ is generated in segments; within a
segment, observations are exchangeable draws from a Normal distribution with
segment-specific parameters, and a new segment begins at each period with
hazard $h$.}

\medskip Assumption A is the detection model, and it is a model of how firms
budget rather than of how markets respond: spending holds at a planned level,
moves abruptly at a planning boundary, and holds again. The
Normal-within-segment choice is a working model for the predictive density;
detection operates on weekly aggregates, where central-limit smoothing makes
it serviceable. Nothing downstream depends on the detector being exact:
condition C2 of Section~\ref{sec:impact} discards detected breaks too small
to matter, so departures from Assumption A cost computation, not validity.

\medskip\noindent\textbf{Assumption B} (Conditional Exogeneity).
\textit{Conditional on $z_t$, the intervention indicator is independent of
the potential outcome path $y_t(0)$ over the pre- and post-windows.}

\medskip Assumption B is the identification core and it is demanding: it
requires that, given the co-observed series, the timing of the intervention
carries no information about how the outcome would have evolved anyway.
Budget changes made in response to demand violate it, in two recognizable
ways. Spending contractually linked to the outcome, as with commission-paid
affiliate channels, produces interventions that follow sales; the resulting
estimates acquire directional significance with the wrong sign, which
condition C4 detects. Budgeting that anticipates demand produces returns
whose magnitude is economically impossible; condition C5 detects that
signature, and Section~\ref{sec:sim} exhibits it in simulation. The demand
covariate in $z_t$ is the observable that carries the burden of conditioning.

\medskip\noindent\textbf{Assumption C} (No Anticipation). \textit{Over the
pre-window, the outcome is generated by the no-intervention regime.}

\medskip Assumption C requires a clean pre-window. Anticipation, as when a
launch is preceded by a ramp of preparatory activity, contaminates its tail.
Two features police it: the dating refinement of Section~\ref{sec:detect}
moves the estimated date to the sharpest daily shift, shortening any
contaminated overlap, and contamination that survives degrades the pre-window
one-step fit that condition C1 requires, so the affected test fails
qualification rather than reporting a biased estimate with confidence.

\medskip\noindent\textbf{Assumption D} (Control Invariance). \textit{The
intervention does not affect $z_t$ over the post-window.}

\medskip Assumption D fails when budgets are reallocated across channels
simultaneously. Condition C3 screens for exactly this, excluding entangled
interventions unless the focal standardized shift dominates by a factor of
three.

\section{Stage One: Detecting and Dating Interventions}\label{sec:detect}
\subsection{The run-length process}
Detection is framed as exact Bayesian inference on the run length of each
activity series: $r_t$ denotes the number of periods since the last
structural break in $x_{k,t}$. Under Assumption A the run length evolves as a
Markov process that either grows by one or resets to zero,
\begin{equation}\label{eq:trans}
P( r_t = r + 1 \mid r_{t-1} = r ) = 1 - h,
\qquad P( r_t = 0 \mid r_{t-1} = r ) = h,
\end{equation}
with $h$ the per-period break hazard. We set $h = 1/13$ per week, one
expected break per planning quarter, matching the cadence at which budgets
are actually revised. The object of inference is the posterior $P(r_t \mid
x_{1:t})$: if most of its mass sits on short run lengths, the data say a
break happened recently, and the location of that mass dates it.

\subsection{The exact filtering recursion}
The posterior is available in closed form by a message-passing argument
\citep{adams07}. Write the joint density of the run length and the data as
$\gamma_t(r) = P(r_t = r, x_{1:t})$. Conditioning on $r_{t-1}$ and applying
the chain rule,
\begin{equation}\label{eq:recur}
\gamma_t(r_t) \;=\; \sum_{r} P( r_t \mid r_{t-1} = r ) \cdot
\pi( x_t \mid r_{t-1} = r, x_{1:t-1} ) \cdot \gamma_{t-1}(r),
\end{equation}
where $\pi( x_t \mid r, x_{1:t-1} )$ is the predictive density of the next
observation given a run of length $r$, which depends on the data only through
the observations of the current segment. The transition kernel
\eqref{eq:trans} has only two nonzero entries per predecessor, so
\eqref{eq:recur} splits into two branch types. Growth branches, $r_t = r +
1$, carry a single predecessor each: $\gamma_t(r+1) = (1 - h) \cdot \pi(x_t
\mid r) \cdot \gamma_{t-1}(r)$. The reset branch, $r_t = 0$, sums over every
predecessor: $\gamma_t(0) = h \cdot \pi_0(x_t) \cdot \sum_r \gamma_{t-1}(r)$,
where $\pi_0$ is the predictive under the prior, because conditional on a
break $x_t$ is the first draw of a new segment. Normalizing $\gamma_t$ across
run lengths gives $P(r_t \mid x_{1:t})$ exactly; no approximation is
involved. The filter maintains $t$ states at time $t$, so a full pass costs
$O(T^2)$ predictive evaluations, which is negligible at the weekly and daily
scales of marketing data.

\medskip\noindent\textbf{Remark 1} (The reset branch). The distinction
between $\pi$ and $\pi_0$ is not cosmetic. If the reset branch reuses the
run's predictive $\pi$ instead of the prior predictive $\pi_0$, both branches
assign $x_t$ identical likelihood, the posterior reset probability collapses
to the hazard $h$ at every period regardless of the data, and the filter
never detects anything. The error is easy to make and announces itself only
as an empty candidate set.

\subsection{Conjugate predictive densities}
Within a segment, observations are Normal with unknown mean and variance, and
the Normal-Inverse-Gamma prior $(\mu_0, \kappa_0, \alpha_0, \beta_0)$ is
conjugate: after the observations of a run, the posterior is
Normal-Inverse-Gamma with parameters $(\mu_r, \kappa_r, \alpha_r, \beta_r)$
obtained by the one-observation recursions
\begin{equation}\label{eq:nig}
\mu' = \frac{\kappa\mu + x_t}{\kappa + 1}, \quad
\kappa' = \kappa + 1, \quad
\alpha' = \alpha + \tfrac{1}{2}, \quad
\beta' = \beta + \frac{\kappa (x_t - \mu)^2}{2(\kappa + 1)},
\end{equation}
applied along each surviving run. Integrating the Normal likelihood over the
Normal-Inverse-Gamma posterior gives the predictive in closed form as a
Student-$t$ density,
\begin{equation}\label{eq:pred}
\pi( x_t \mid r ) \;=\;
t\!\left( x_t ;\; 2\alpha_r ,\; \mu_r ,\;
\frac{\beta_r (\kappa_r + 1)}{\alpha_r \kappa_r} \right),
\end{equation}
with $2\alpha_r$ degrees of freedom, location $\mu_r$, and squared scale
$\beta_r(\kappa_r + 1)/(\alpha_r \kappa_r)$. The heavy tails of the
Student-$t$ at small run lengths are what make the filter usable on real
budget data: a single outlying week inside a stable segment raises the reset
probability only transiently, because the wide predictive of the young
competing run explains the outlier almost as well as a genuine break would.

\subsection{Declaration, dating, and refinement}
Operational interventions phase in rather than arriving between two adjacent
observations: a budget decision made on a Friday executes across a week
boundary. A single-step rule based on $P(r_t = 0 \mid x_{1:t})$ misses such
breaks. Declaration therefore accumulates evidence over short run lengths: a
break is declared at the first $t$ with
\begin{equation}\label{eq:declare}
P( r_t \le r^{*} \mid x_{1:t} ) \;>\; \lambda,
\qquad r^{*} = 4, \quad \lambda = 0.6,
\end{equation}
dated $\hat t^{*} = t - \hat r$ at the modal short run length, and refined on
daily data to the day that maximizes the shift between adjacent 14-day means.
Detection runs on weekly aggregates; the refinement recovers the daily date.

\medskip\noindent\textbf{Remark 2} (Detection is allowed to be liberal). The
constants $(h, r^{*}, \lambda)$ trade recall against false discovery, and we
resolve the trade in favor of recall deliberately: the qualification stage of
Section~\ref{sec:impact}, not the detector, carries the burden of validity,
and a spurious candidate costs a rejected test rather than a wrong estimate.
Appendix~\ref{app:constants} collects all constants with their roles.

\section{Stage Two: The Counterfactual Model}\label{sec:cf}
\subsection{A structural time series model for the outcome}
Detection hands over a candidate pair $(k, \hat t^{*})$. Estimation now moves
to the outcome variable. Fix a pre-window of $P$ periods ending at $\hat
t^{*}$. The counterfactual model for the outcome is the structural time
series form \citep{harveytodd83}
\begin{equation}\label{eq:model}
\log y_t = \mu_t + z_t'\beta + \varepsilon_t, \qquad
\mu_t = \mu_{t-1} + \delta_{t-1} + \eta_t, \qquad
\delta_t = \delta_{t-1} + \zeta_t,
\end{equation}
with $\varepsilon_t \sim N(0, \sigma_\varepsilon^2)$, $\eta_t \sim N(0,
\sigma_\eta^2)$, $\zeta_t \sim N(0, \sigma_\zeta^2)$, and $z_t$ the
co-observed information set of Section~\ref{sec:setup}. The unobserved level
$\mu_t$ and slope $\delta_t$ absorb evolution the regressors cannot
represent; the regression term carries seasonality, the other channels, and
demand. The use of many co-observed series as the forecast information base
parallels diffusion-index forecasting \citep{stockwatson02}. In state space
form, with state vector $\alpha_t = (\mu_t, \delta_t)'$,
\begin{equation}\label{eq:ss}
\log y_t = z_t'\beta + H \alpha_t + \varepsilon_t, \qquad
\alpha_t = T \alpha_{t-1} + u_t, \qquad
H = (1, 0), \quad
T = \begin{bmatrix} 1 & 1 \\ 0 & 1 \end{bmatrix}, \quad
u_t \sim N(0, Q),
\end{equation}
with $Q = \mathrm{diag}(\sigma_\eta^2, \sigma_\zeta^2)$.

\subsection{Filtering, likelihood, and estimation}
For given parameters the Kalman filter computes the model exactly. With
$a_{t|t-1}$ and $P_{t|t-1}$ the one-step state mean and variance, the
recursion alternates prediction and update,
\begin{equation}\label{eq:kfpred}
a_{t|t-1} = T a_{t-1}, \qquad
P_{t|t-1} = T P_{t-1} T' + Q; \qquad
\nu_t = \log y_t - z_t'\beta - H a_{t|t-1}, \qquad
F_t = H P_{t|t-1} H' + \sigma_\varepsilon^2,
\end{equation}
\begin{equation}\label{eq:kfup}
a_t = a_{t|t-1} + P_{t|t-1} H' F_t^{-1} \nu_t, \qquad
P_t = P_{t|t-1} - P_{t|t-1} H' F_t^{-1} H P_{t|t-1},
\end{equation}
initialized diffusely, and the innovations deliver the likelihood by the
prediction error decomposition,
\begin{equation}\label{eq:lik}
\log L( \beta, \sigma_\varepsilon ; \sigma_\eta, \sigma_\zeta )
\;=\; -\frac{1}{2} \sum_{t > d} \left[ \log 2\pi F_t + \frac{\nu_t^2}{F_t}
\right],
\end{equation}
with the first $d = 2$ innovations of the diffuse initialization dropped from
the sum.

The free parameters $(\beta, \sigma_\varepsilon)$ are estimated by maximizing
\eqref{eq:lik} over the pre-window. The state variances are not estimated;
they are fixed by the identification constraint that defines the estimator,
\begin{equation}\label{eq:constraint}
\sigma_\eta = c_\eta \cdot \sd( \log y_{\mathrm{pre}} ), \qquad
\sigma_\zeta = c_\zeta \cdot \sd( \log y_{\mathrm{pre}} ), \qquad
c_\eta = 0.01, \quad c_\zeta = 0.001.
\end{equation}

\subsection{Why the constraint: identification, not convenience}
The local level in \eqref{eq:model} and the regression term $z_t'\beta$
compete for the same variation. If $(\sigma_\eta, \sigma_\zeta)$ are
estimated freely, the smoothed level absorbs the regression signal in-sample,
the fitted model looks excellent, and the post-window projection carries a
predictive variance that grows without information, because a level that
explained everything in-sample forecasts nothing out of sample.
Section~\ref{sec:bench} quantifies the consequence on the benchmark: the
free-level variant's interval on the positive experiment is 2.2 times as
wide as the constrained estimator's and reaches from one percent of the
experimental estimate to 1.8 times it. The constraint \eqref{eq:constraint}
pins the level to movements of about one percent of pre-window outcome
dispersion per step: small enough that the regressors must do the explanatory
work, large enough to absorb slow drift the regressors cannot represent. The
claim embedded in \eqref{eq:constraint}, that the outcome's slow evolution is
carried by observable demand rather than by an unobservable random walk, is
substantive, and it is checkable: when it is false the constrained model
cannot track the pre-window, and condition C1 fails the test rather than
letting it report a confident error.

\medskip\noindent\textbf{Remark 3} (Sensitivity). Across $c_\eta \in \{0.01,
0.03, 0.05\}$ the point estimate $\hat\tau$ is stable in the applications of
Sections~\ref{sec:bench} and \ref{sec:sim}, so conclusions do not hinge on
the exact value of the constant. Estimating the state variances by maximum
likelihood is not a viable middle course: the estimated variances drift
upward and the intervals degenerate toward the free-level behavior documented
in Section~\ref{sec:bench}.

\medskip Two window choices complete the specification. The pre-window is $P
= 52$ weeks on weekly data, and on daily data the available pre-treatment
days up to roughly one quarter of a year of history per regressor examined; a
26-week alternative was rejected because regression coefficients become
unstable across adjacent candidate dates. The post-window is $W = 8$ weeks,
or the documented intervention length on daily data: a longer window adds
spending change roughly linearly while compounding the extrapolation drift
any counterfactual accumulates. Where the candidate set permits, the
pre-window should contain no focal activity, launches and restarts rather
than pauses, for the reason formalized next.

\medskip\noindent\textbf{Proposition 1} (Donor contamination). \textit{Suppose
the focal channel is active in the pre-window with effect $\theta$ on $\log
y$. Let $\Omega$ denote the pre-window covariance of $H\alpha_t +
\varepsilon_t$ implied by \eqref{eq:constraint}, and $\Gamma =
(Z'\Omega^{-1}Z)^{-1} Z'\Omega^{-1} x_k$ the corresponding projection
coefficient vector of $x_k$ on $z$. Then, conditional on the regressors, the
constrained estimator satisfies $\mathrm{E}[\hat\beta] = \beta +
\Gamma\theta$, and the counterfactual inherits the bias $z_t'\Gamma\theta$
over the post-window. If $x_k = 0$ throughout the pre-window, the
contamination vanishes.} (Proof sketch in Appendix~\ref{app:proof}.)

\medskip\noindent\textbf{Corollary 1} (Design rule). \textit{Among candidate
interventions for a channel, prefer those whose pre-window contains no focal
activity.} In the simulations of Section~\ref{sec:sim} the mechanism is
quantitatively visible: pause-type tests with active pre-windows overshoot
known truth by a factor near two, while the matching focal-free restart errs
by 11 percent.

\section{The Causal Impact Analysis}\label{sec:impact}
\subsection{Simulation-based estimation and inference}
The fitted model is projected over the post-window with observed regressor
paths, and $M = 2{,}000$ counterfactual paths $\tilde y(m)$ are simulated
with both state and measurement noise, exponentiated back to the outcome
scale. The estimator, its standard error, and directional significance are
\begin{equation}\label{eq:tau}
\hat\tau = \sum_{\mathrm{post}} y_t
- \frac{1}{M} \sum_{m} \sum_{\mathrm{post}} \tilde y_t(m),
\qquad
\se(\hat\tau) = \sd_m\!\left[ \sum_{\mathrm{post}} y_t
- \sum_{\mathrm{post}} \tilde y_t(m) \right],
\end{equation}
\begin{equation}\label{eq:pdir}
\pdir = \frac{1}{M} \sum_{m}
\mathbf{1}\!\left\{ \sgn\!\Bigl( \sum_{\mathrm{post}} y_t
- \sum_{\mathrm{post}} \tilde y_t(m) \Bigr) = \sgn(\Delta s) \right\},
\end{equation}
with 90 percent intervals from empirical quantiles of the simulated
cumulative differences. Serial dependence of imputation errors is carried by
the path distribution rather than by per-period bands, in line with the
counterfactual-imputation literature \citep{masini22,goncalvesng24}. The
simulated paths condition on the estimated $(\beta, \sigma_\varepsilon)$;
parameter uncertainty is not propagated, so the intervals are prediction
intervals given the fitted model, and their adequacy is checked rather than
assumed, in-sample by the calibration half of C1 and out of sample by the
benchmark coverage of Section~\ref{sec:bench}. The counterfactual is a
forecast and is diagnosed as one: pre-window one-step-ahead residuals must
attain $R^2 \ge 0.7$, and their probability integral transform (PIT) must not
reject uniformity at Kolmogorov-Smirnov (KS) $p \ge 0.01$, so that the
simulated dispersion used for inference is neither optimistic nor padded
\citep{diebold95,harveykoopman92,venkatraman24}; directional evaluation
follows \citet{pesaran92}.

\subsection{Qualification}
\begin{table}[t]
\centering\small
\caption{Qualification conditions.}\label{tab:qual}
\begin{tabular}{p{0.16\textwidth} p{0.38\textwidth} p{0.36\textwidth}}
\toprule
Condition & Requirement & Failure mode excluded \\
\midrule
C1 Adequacy & one-step $R^2 \ge 0.7$; PIT uniformity, KS $p \ge 0.01$ &
extrapolating a model that never fit \\
C2 Materiality & $|\Delta \text{activity}| \ge 20\%$ of pre level, or a
launch & reading noise as treatment \\
C3 Isolation & no concurrent break without $3\times$ dominant standardized
shift & entangled interventions (Assumption D) \\
C4 Direction & $\pdir \ge 0.95$ & outcome-linked spending (Assumption B) \\
C5 Economics & $\hat\rho$ and $\rho_{\min}$ within the plausible range for
the channel's scale & demand-following budgets (Assumption B); untestable
designs \\
\bottomrule
\end{tabular}
\end{table}
Table~\ref{tab:qual} collects the five conditions, ordered from the
statistical to the economic. C1 asks whether the counterfactual ever fit, and
checks calibration as well as accuracy, since an interval from a
miscalibrated forecast is decoration. C2 imposes materiality, because small
interventions produce estimates dominated by imputation noise. C3 imposes
isolation, as discussed under Assumption D. C4 requires directional
significance at 0.95, and an estimate that is significant with the wrong sign
is not an unlucky draw but the signature of outcome-linked spending. C5 asks
whether the implied economics are possible; Section~\ref{sec:sim} shows a
spurious candidate passing every statistical screen while failing exactly
this one.

\subsection{Power}
The power bound is closed form: the smallest return the design could declare
significant at level $\alpha$ is
\begin{equation}\label{eq:rhomin}
\rho_{\min} \;=\; \frac{z_{1-\alpha} \cdot \se(\hat\tau)}{|\Delta s|},
\end{equation}
and guaranteeing power $1 - \beta$ against a given alternative multiplies the
bound by $(z_{1-\alpha} + z_{1-\beta}) / z_{1-\alpha}$. When $\rho_{\min}$
exceeds any economically possible return, the channel is untestable at that
aggregation: no estimate the design could produce would be worth reading.
This is formula \eqref{eq:power} of the introduction made operational, and it
delivers the \citet{lewisrao15} power verdict for a specific channel,
dataset, and window before resources are committed rather than after.

\begin{algorithm}[H]
\small
\begin{singlespace}
\caption{Pseudo-incrementality testing.}\label{alg:pit}
\begin{algorithmic}[1]
\Require outcome $y_t$, $t = 1, \ldots, T$; activities $x_{k,t}$ and spending
$s_{k,t}$, $k = 1, \ldots, K$; demand covariates $d_t$; the constants of
Appendix~\ref{app:constants}: hazard $h$, declaration pair $(r^{*},
\lambda)$, variance constants $(c_\eta, c_\zeta)$, windows $(P, W)$,
simulation size $M$
\Ensure qualified lift estimates with intervals and returns; documented
refusals; power verdicts
\Statex \textbf{Stage 1: detection and dating (Section~\ref{sec:detect})}
\For{$k = 1$ \textbf{to} $K$}
  \State aggregate $x_{k,t}$ to weekly totals; initialize the run-length
  filter with the Normal-Inverse-Gamma prior
  \For{$t = 1$ \textbf{to} $T$}
    \State update $\gamma_t(r)$ by the growth and reset branches of
    \eqref{eq:recur}, with predictives \eqref{eq:nig} and \eqref{eq:pred}
    \If{$P( r_t \le r^{*} \mid x_{1:t} ) > \lambda$}
      \State date $\hat t^{*} = t - \hat r$ at the modal short run; refine on
      daily data; add $(k, \hat t^{*})$ to the candidate set $\mathcal{C}$
    \EndIf
  \EndFor
\EndFor
\Statex \textbf{Stage 2: causal impact analysis (Sections~\ref{sec:cf} and
\ref{sec:impact})}
\For{\textbf{each} $(k, \hat t^{*}) \in \mathcal{C}$}
  \State build $z_t$ from the co-observed set: remaining channels, calendar
  terms, annual lag, demand covariate
  \State fix the pre-window of $P$ periods ending at $\hat t^{*}$; set
  $(\sigma_\eta, \sigma_\zeta)$ by \eqref{eq:constraint}; estimate $(\beta,
  \sigma_\varepsilon)$ by maximizing \eqref{eq:lik}
  \State simulate $M$ counterfactual paths over the $W$ post-window periods;
  compute $\hat\tau$, $\se(\hat\tau)$, the 90\% interval, and $\pdir$ by
  \eqref{eq:tau} and \eqref{eq:pdir}; set $\hat\rho = \hat\tau / \Delta s$
  \State evaluate conditions C1 through C5 (Table~\ref{tab:qual})
  \If{all five conditions pass}
    \State report $(\hat\tau, \hat\rho)$, the interval, and $\pdir$ as
    experimental-grade evidence
  \Else
    \State report the failed condition and its failure signature
  \EndIf
  \State compute $\rho_{\min}$ by \eqref{eq:rhomin}; \textbf{if}
  $\rho_{\min}$ exceeds any plausible return for the channel, report the
  channel untestable at this aggregation
\EndFor
\end{algorithmic}
\end{singlespace}
\end{algorithm}

\section{Relation to Existing Methods}\label{sec:relation}
\begin{table}[t]
\centering\small
\caption{Capability comparison.}\label{tab:capab}
\begin{tabular}{p{0.30\textwidth} p{0.12\textwidth} p{0.20\textwidth}
p{0.11\textwidth} p{0.13\textwidth}}
\toprule
Method & Date known? & Data required & Interval? & Discovers
interventions? \\
\midrule
Intervention analysis \citep{boxtiao75} & yes & single series & yes & no \\
Causal impact / BSTS, free level \citep{brodersen15} & yes & outcome +
controls & yes (wide) & no \\
Synthetic control \citep{abadie21} & yes & panel of untreated units &
placebo-based & no \\
Difference-in-differences & yes & panel of untreated units & yes & no \\
Geo-experiment analysis \citep{vaver11} & yes (designed) & randomized geo
panel & yes & no \\
Proposed procedure & no (estimated) & outcome + co-observed series & yes &
yes \\
\bottomrule
\end{tabular}
\par\smallskip
{\footnotesize\itshape ``Interval?'' refers to routinely available
uncertainty statements for the cumulative effect of a single intervention on
a single treated series.}
\end{table}
Three distinctions organize Table~\ref{tab:capab}. First, discovery: all
existing methods condition on a known intervention date; in operational data
the dates are unrecorded, plural, and contested, and stage one supplies them
with posterior probabilities attached. Second, data requirements: the panel
methods need untreated units, which exist only for advertisers with
geo-resolved reporting; the proposed procedure requires only series
co-observed with the outcome. Third, the free-level issue: causal impact
estimation with an unconstrained local level, the Bayesian structural time
series (BSTS) default, is the closest relative of stage two, and
Section~\ref{sec:bench} shows what the difference, exactly the constraint
\eqref{eq:constraint}, does to interval width. A fourth relation is to the
forecasting-evaluation literature: the procedure treats causal estimation as
counterfactual forecasting, so its diagnostics are forecasting diagnostics
\citep{diebold95,pesaran92,perron21}, and the counterfactual-imputation
results of \citet{masini22} and \citet{goncalvesng24} shape the
path-distribution inference in \eqref{eq:tau} and \eqref{eq:pdir}.

\section{Validation on Meta's Published Experiments}\label{sec:bench}
\subsection{The data}
No public dataset pairs a real advertiser's lift study with its underlying
time series, so the strongest available external test is the benchmark Meta
published with its GeoLift experimentation framework \citep{meta22}: two
city-level experiments with documented designs and published experimental
estimates, released precisely so that third parties can reproduce the
analysis from the files. We use these two datasets and nothing else; both are
simulated by their publisher, and they are nonetheless the correct comparison
targets, because they are the platform's own reference analyses of
experiments whose every design detail is verifiable from the released files.

\begin{figure}[t]
\centering
\includegraphics[width=\textwidth]{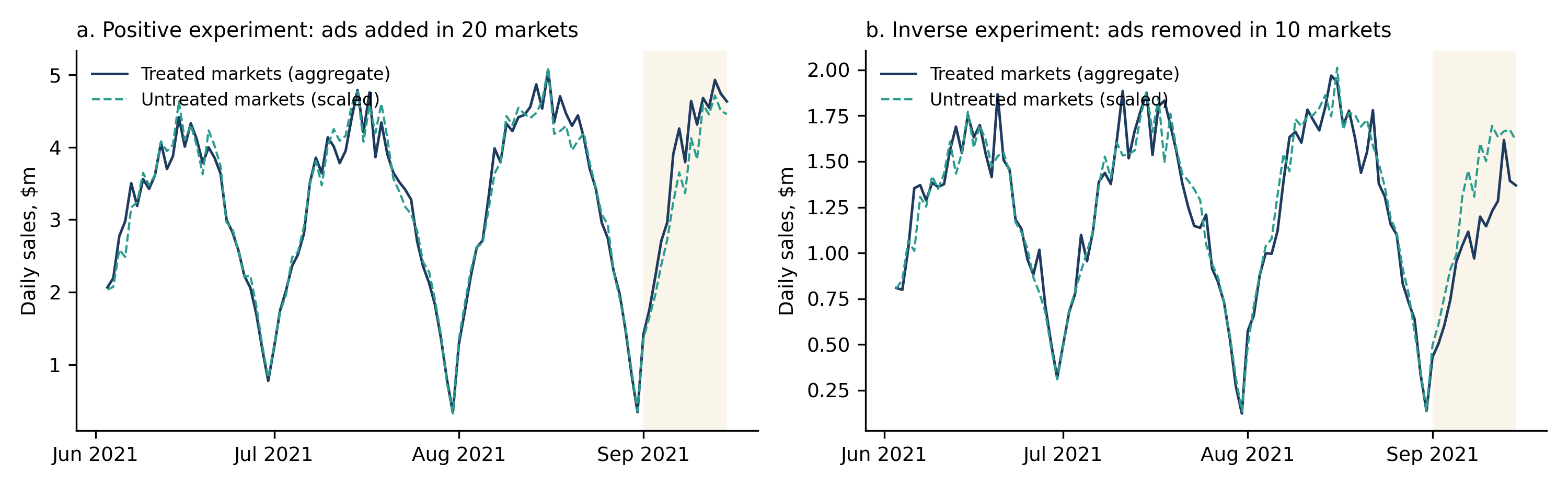}
\caption{The two datasets: treated-market aggregate daily sales (navy)
against the untreated-market aggregate scaled to the treated pre-period mean
(teal), with the September 1-15 test window shaded.}\label{fig:data}
\end{figure}

Each dataset is a panel of daily sales for 40 United States cities over the
105 days from June 3 to September 15, 2021, with a geographic test on
September 1 through 15. In the positive experiment, advertising was added in
20 documented markets; GeoLift's augmented synthetic control analysis puts
the campaign's effect at $+$7,159,530 incremental sales. In the inverse
experiment, advertising was removed in 10 documented markets while 30 markets
continued spending on a released daily schedule totaling 6,884,784; the
published analysis concludes the dark markets sacrificed 3,366,433 in sales
against the 2,104,021 of spending their synthetic counterpart received, an
incremental return of 1.60 per unit of spending. Table~\ref{tab:data}
collects the facts; Figure~\ref{fig:data} shows the series. Two features
matter for what follows. The dominant variation is a pronounced weekly cycle,
with intra-week swings of several multiples of the trough, which any
counterfactual must absorb before treatment effects of 15 percent are
visible. And the treated and untreated aggregates track each other closely
for 90 days and then separate on September 1, in opposite directions in the
two experiments, which is the entire effect to be measured.

\begin{table}[t]
\centering\small
\caption{The two Meta GeoLift benchmark datasets.}\label{tab:data}
\begin{tabular}{p{0.34\textwidth} p{0.28\textwidth} p{0.28\textwidth}}
\toprule
 & Positive experiment & Inverse experiment \\
\midrule
Panel & 40 cities $\times$ 105 days & 40 cities $\times$ 105 days \\
Sample & June 3 to September 15, 2021 & June 3 to September 15, 2021 \\
Test window & September 1 to 15 (15 days) & September 1 to 15 (15 days) \\
Intervention & ads added in 20 markets & ads removed in 10 markets \\
Design target effect size (power analysis) & $+$4 percent & $-$2 percent \\
Documented spending & not released & 6,884,784 over 30 spending markets \\
Synthetic-counterpart spending, dark markets & - & 2,104,021 \\
Published estimate & $+$7,159,530 & $-$3,366,433 \\
Published return & not released & 1.60 per unit of spending \\
\bottomrule
\end{tabular}
\par\smallskip
{\footnotesize\itshape All entries are taken from the released data files and
the accompanying whitepaper \citep{meta22}; the whitepaper's totals reproduce
from the files. Design target effect sizes are the minimum detectable effects
used for market selection; the simulated effects, as published, are larger.
Synthetic-counterpart spending is the whitepaper's synthetic-control-weighted
investment, the denominator of the published return.}
\end{table}

\subsection{How the data are treated to validate the model}
The validation protocol is fixed before any estimate is computed, and every
tuning constant is the one stated in Sections~\ref{sec:detect} through
\ref{sec:impact}; nothing is adapted to the benchmark. For each experiment,
the focal outcome series is the aggregate of the documented treated markets,
and the demand covariate is the aggregate of the untreated markets, the
construction Section~\ref{sec:setup} recommends whenever outcomes are
recorded by region. The procedure therefore sees two series and a spending
schedule; it uses neither the panel's cross-sectional structure, nor the
randomization, nor the published estimates. Stage one runs the detector of
Section~\ref{sec:detect} on the released spending schedule to date the
intervention. Stage two fits the constrained model of Section~\ref{sec:cf}
on the 90 pre-treatment days, with day-of-week terms and the log untreated
aggregate as $z_t$, and projects it over the 15 test days. The inference step
of stage two then computes \eqref{eq:tau} and \eqref{eq:pdir} from 2,000
simulated paths and evaluates the qualification conditions. Only then are the
results compared with the published experimental estimates. Success is
defined in advance: the intervention should be found and dated, the sign
should be right with $\pdir \ge 0.95$, the 90 percent interval should cover
the experimental estimate and exclude zero, and the point estimate should be
close enough to the experimental one to support the same budget decision.

\subsection{Stage-one results: finding the intervention}
\begin{figure}[t]
\centering
\includegraphics[width=\textwidth]{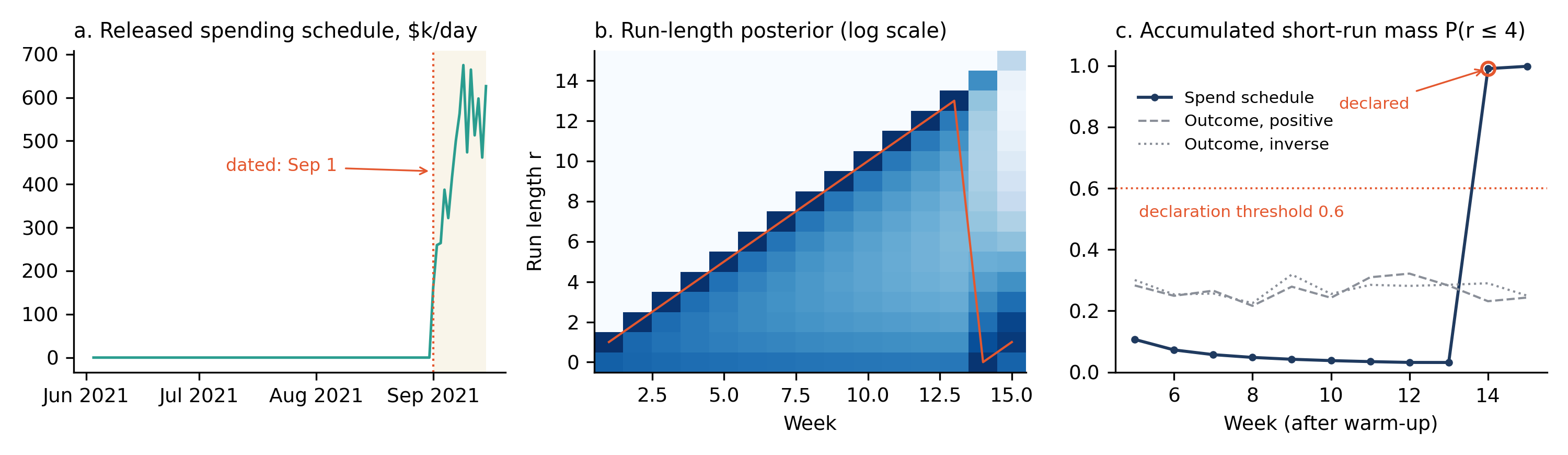}
\caption{Stage one on the benchmark: (a) the released spending schedule with
the refined date; (b) the run-length posterior collapsing at the break; (c)
accumulated short-run mass against the declaration threshold, for the
spending schedule and, for contrast, the two outcome
series.}\label{fig:detect}
\end{figure}

The only activity series released with the benchmark is the daily spending
schedule of the markets that advertised through the test window, and its
September 1 start is documented rather than hidden, so running stage one on
it is a check of the dating machinery, not a discovery test; the discovery
test, on series whose breaks the procedure has never seen documented, is the
simulation study of Section~\ref{sec:sim}. The machinery passes cleanly. At
weekly aggregation the accumulated short-run mass crosses the declaration
threshold in the first full week of spending, reaching 0.99, and the daily
refinement dates the intervention to September 1 exactly.
Figure~\ref{fig:detect} shows it at work: the run-length posterior climbs the
diagonal for thirteen weeks, collapses to zero at the break, and the
short-run mass jumps from under 0.05 to 0.99. Applied instead to the outcome
series, the same filter flags nothing: fifteen weekly observations leave the
maximum accumulated mass at 0.32, well below the threshold. The contrast
is the design lesson of stage one. Activity series break sharply, because
budgets are decisions; outcomes break noisily, because they are consequences.
Detection belongs on the decision variable, which is also what keeps it free
of selection on outcomes: the procedure never chooses tests by looking at the
variable on which effects are then measured.

\subsection{Stage-two results: the causal impact analysis}
\begin{figure}[t]
\centering
\includegraphics[width=\textwidth]{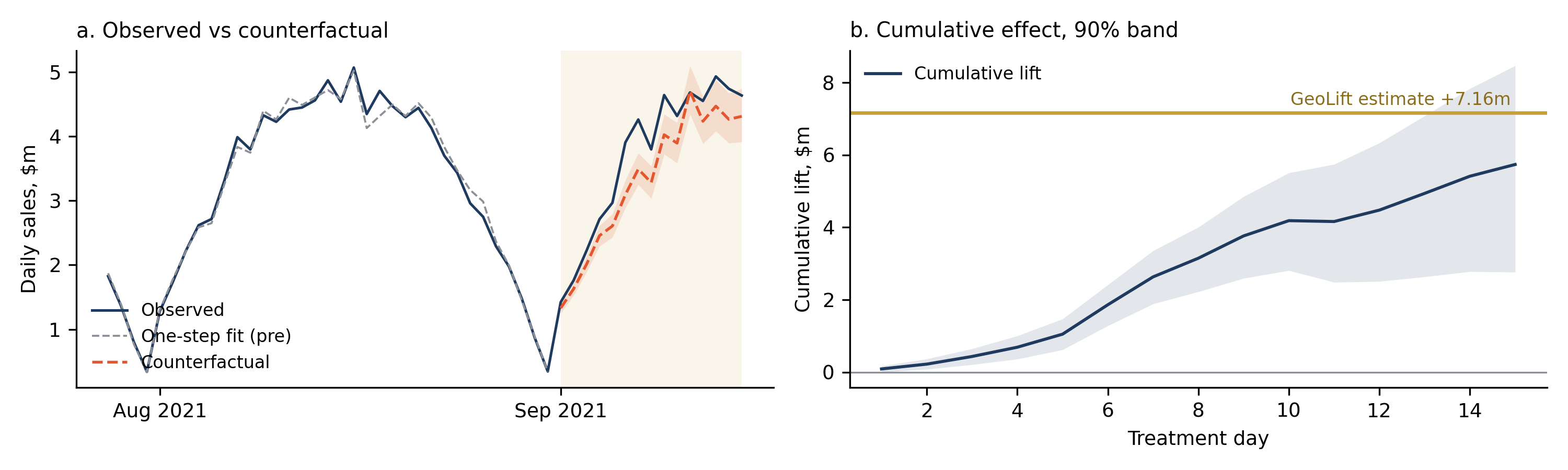}
\caption{Causal impact analysis, positive experiment: (a) observed
treated-market sales, the pre-window one-step fit, and the counterfactual
projection with 90\% band; (b) the cumulative lift path against the published
experimental estimate.}\label{fig:cfpos}
\end{figure}

\begin{figure}[t]
\centering
\includegraphics[width=\textwidth]{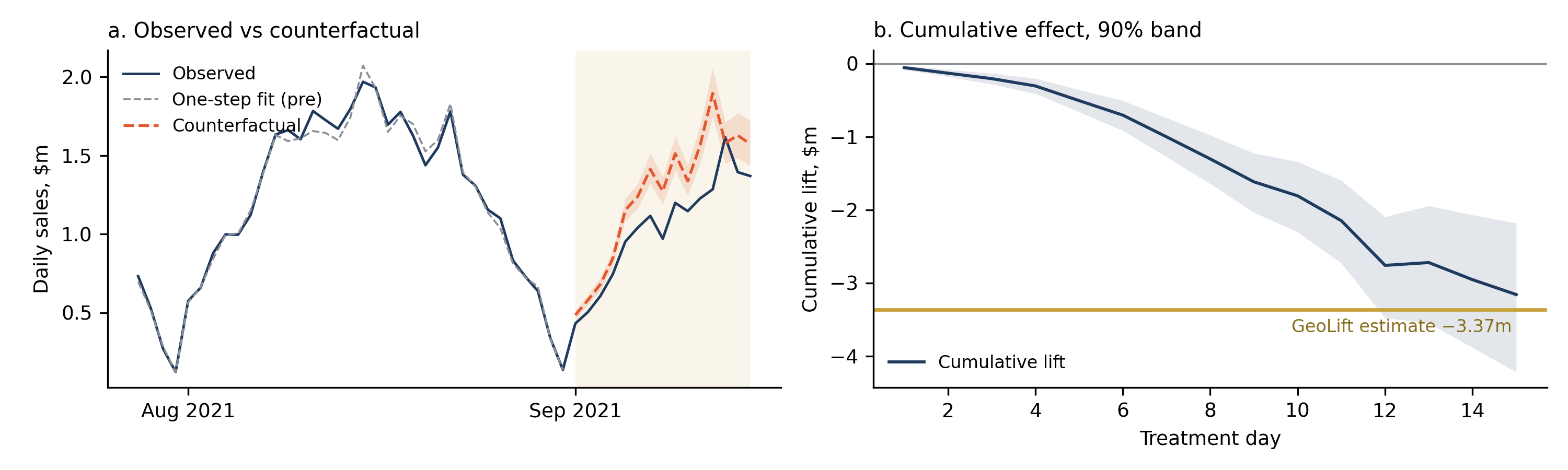}
\caption{Causal impact analysis, inverse experiment: layout as in
Figure~\ref{fig:cfpos}. The estimator tracks the removal of
advertising.}\label{fig:cfneg}
\end{figure}

The constrained model tracks both pre-windows almost perfectly: one-step
$R^2$ is 0.997 and 0.997 against the C1 floor of 0.7, and the PIT of the
one-step residuals accepts uniformity (KS $p =$ 0.12 and 0.89), so C1
passes and the simulated dispersion can be taken at face value. C2 is met by
construction, a launch in one experiment and a full removal in the other; C3
is met by the controlled design, which changes one channel at a time; C4
delivers $\pdir =$ 0.999 for the positive experiment and 1.000 for
the inverse. C5 is evaluable only where spending is documented: on the
inverse experiment the implied return of 1.50 sits inside any plausible
range for the channel, whose historical figure in the whitepaper is 1.6, so
C5 passes; on the positive experiment spending was not released, so C5 cannot
be computed, and the randomized design rules out the demand-following failure
mode C5 exists to catch. Figures~\ref{fig:cfpos} and \ref{fig:cfneg} display
the analyses; Tables~\ref{tab:pos} and \ref{tab:neg} report the horse race
against five alternative estimators computed on the same released data, each
estimator consuming the representation it requires.

\begin{table}[t]
\centering\small
\caption{Positive experiment (published estimate $+$7,159,530).}\label{tab:pos}
\begin{tabular}{l l l l}
\toprule
Estimator & Estimate (error) & 90\% interval & Covers benchmark \\
\midrule
Pre-post & 7,684,575 ($+$7.3\%) & - & - \\
Static OLS & 4,561,919 ($-$36.3\%) & [3,689,530; 5,430,245] & no \\
BSTS (free level) & 6,837,795 ($-$4.5\%) & [66,272; 12,885,430] & yes \\
Constrained UCM (proposed) & 5,737,904 ($-$19.9\%) & [2,764,721; 8,465,851] & yes \\
Difference-in-differences & 4,487,844 ($-$37.3\%) & - & - \\
Synthetic control & 4,003,737 ($-$44.1\%) & - & - \\
\bottomrule
\end{tabular}
\par\smallskip
{\footnotesize\itshape Estimator labels: OLS, ordinary least squares; BSTS,
Bayesian structural time series with freely estimated level; UCM, unobserved
components model, the proposed constrained estimator. The first four rows use
the treated and untreated aggregates only; the panel methods additionally use
the market-level panel.}
\end{table}

\begin{table}[t]
\centering\small
\caption{Inverse experiment (published estimate $-$3,366,433).}\label{tab:neg}
\begin{tabular}{l l l l}
\toprule
Estimator & Estimate (error) & 90\% interval & Covers benchmark \\
\midrule
Pre-post & $-$3,173,832 ($+$5.7\%) & - & - \\
Static OLS & $-$3,686,541 ($-$9.5\%) & [$-$4,043,196; $-$3,330,921] & yes \\
BSTS (free level) & $-$3,134,768 ($+$6.9\%) & [$-$4,644,367; $-$1,661,189] & yes \\
Constrained UCM (proposed) & $-$3,159,030 ($+$6.2\%) & [$-$4,218,615; $-$2,179,382] & yes \\
Difference-in-differences & $-$3,611,035 ($-$7.3\%) & - & - \\
Synthetic control & $-$2,899,619 ($+$13.9\%) & - & - \\
\bottomrule
\end{tabular}
\end{table}

\subsection{Reading the evidence: is this accuracy good?}
A percentage error means nothing without a scale, so we supply three. The
first is the economic scale a manager decides on. The inverse experiment puts
the return to the Meta channel at 1.60 per unit of spending; the proposed
estimator, run blind on the same series, arrives at 1.50. No budget
decision distinguishes 1.50 from 1.60. On the positive experiment the
published estimate is a lift of 14.8 percent of baseline sales and the
estimator says 11.5 percent: the same finding, a large, significant,
profitable launch, stated about a fifth too conservatively.

The second scale is what the alternatives achieve on the same data. On the
positive experiment the static regression, difference-in-differences, and
synthetic control miss by 36 to 44 percent, with the two panel methods
failing because the treated-control relation shifts in September and simple
implementations assume it fixed. The free-level variant is the honest
competitor: its point error is 4.5 percent, better than the proposed
19.9 percent, but its interval spans 66,272 to 12,885,430, from one percent
of the experimental estimate to 1.8 times it, an answer consistent with the
campaign having accomplished almost nothing and with it having doubled the
experiment's result. The proposed interval is 2.2 times narrower and still
covers. On the inverse experiment the proposed estimator errs 6.2 percent
with a covering interval that excludes zero; the static regression's interval
is narrower there and also covers, but the same estimator misses the positive
experiment by 36 percent, which is why Section~\ref{sec:verdict} judges the
estimators jointly across both experiments. The panel rows also carry a
caution against overreading: the released target is itself produced by
GeoLift's augmented synthetic control with bias correction, so the simple
synthetic control row measures what an off-the-shelf implementation achieves
at this aggregation, not the ceiling of the method class.

The third scale is the naive baseline. Pre-post errs only 7.3 and 5.7
percent here, and that number is a warning rather than a recommendation: it
reflects a flat baseline over a short window, offers no uncertainty
statement, and Section~\ref{sec:sim} shows the same estimator missing by
factors of three to five, and getting the sign wrong, the moment baselines
trend.

\section{Operating Characteristics in Simulation}\label{sec:sim}
Two experiments cannot map failure modes, so we chart the procedure's
operating characteristics where truth is manipulable. Five three-year weekly
markets are generated with an open-source simulator \citep{quach26} following
the siMMMulator process \citep{facebook22}: four to six paid channels,
geometric adstock, concave response, weather-driven seasonality, trend, and
one designed intervention per market, two launches, two sustained cuts, one
sustained increase, with noise tuned so a well-specified regression attains
$R^2$ between 0.90 and 0.95. Run blind on all five markets, the procedure
disposes of the five designed interventions as follows. Three, a launch, the
increase, and one cut, were detected on the designed channel, matched to the
designed intervention, and issued qualified estimates. The second launch was
detected but refused under C4: directional significance reached only $\pdir =
0.57$ against the required 0.95, so no estimate was issued; the launch was
real, its effect was not separable from noise, and the procedure said so
rather than guessing. The second cut was not recovered: in its market the
detector instead surfaced a spurious candidate on a different channel, whose
estimated return of 10.3 exceeds by two orders of magnitude the simulator's
true return of 0.07 for that channel, exactly the economic impossibility that
C5 rejects. One designed intervention missed, one refused, and one spurious
candidate caught by economics rather than statistics: an instrument for
finding effects in observational data earns trust by what it refuses.

\begin{table}[t]
\centering\footnotesize
\caption{Estimator comparison on the simulated markets (selected
tests).}\label{tab:sim}
\begin{tabular}{l l l l l l}
\toprule
Design & True effect & Estimator & Estimate & 90\% interval & Covers truth \\
\midrule
Launch & 1,151,635 & Pre-post & 3,811,893 & - & - \\
 &  & Static OLS & 2,044,741 & [1,243,494; 2,812,083] & no \\
 &  & BSTS (free level) & 2,212,355 & [$-$207,893; 4,495,924] & yes \\
 &  & Constrained UCM (proposed) & 2,393,836 & [1,235,230; 3,553,883] & no \\
\midrule
Cut (spurious cand.) & 14,417 & Pre-post & $-$3,221,176 & - & - \\
 &  & Static OLS & 2,409,245 & [1,995,617; 2,813,246] & no \\
 &  & BSTS (free level) & 1,543,951 & [986,683; 2,105,642] & no \\
 &  & Constrained UCM (proposed) & 1,525,393 & [898,960; 2,130,135] & no \\
\midrule
Increase & 1,155,320 & Pre-post & 5,262,508 & - & - \\
 &  & Static OLS & 2,201,464 & [1,581,182; 2,798,981] & no \\
 &  & BSTS (free level) & 3,150,366 & [2,327,786; 3,960,475] & no \\
 &  & Constrained UCM (proposed) & 3,153,094 & [2,261,590; 3,998,953] & no \\
\midrule
Cut & $-$3,061,716 & Pre-post & 737,328 & - & - \\
 &  & Static OLS & $-$2,424,199 & [$-$3,430,090; $-$1,458,680] & yes \\
 &  & BSTS (free level) & $-$1,321,491 & [$-$3,979,408; 1,147,417] & yes \\
 &  & Constrained UCM (proposed) & $-$1,762,563 & [$-$3,191,494; $-$389,515] & yes \\
\bottomrule
\end{tabular}
\par\smallskip
{\footnotesize\itshape Panel methods are infeasible: the designs contain no
cross-sectional units. The spurious S3 candidate has true effect near zero
for the surfaced channel.}
\end{table}

The interval record must be stated plainly: of the three issued estimates,
the nominal 90 percent interval covered truth only for the cut.
Table~\ref{tab:sim} carries the two findings behind that record, and both
discipline the reading of Section~\ref{sec:bench}. First, with weak controls,
co-observed channels and weather only, all regression-based estimators move
together: they overestimate the launch and the increase by factors near two
to three, and they are all deceived by the spurious candidate. The binding
constraint is the information in the control set, not the estimator applied
to it, which is why the benchmark's untreated-market covariate matters so
much and why Section~\ref{sec:setup} elevates the demand covariate to a
requirement. Second, the naive pre-post estimator that looked adequate on the
benchmark misses here by 231 to 355 percent on the launch and the increase
and gets the sign of the cut wrong, confirming that its
Section~\ref{sec:bench} performance is a property of that dataset's flat
baseline, not of the method. On the cut, the proposed interval covers the
truth and excludes zero; the free-level interval covers only by stretching
across zero.

\section{Verdict: Does the Proposed Procedure Beat the
Alternatives?}\label{sec:verdict}
\begin{table}[t]
\centering\footnotesize
\caption{Verdict criteria on the Meta benchmark.}\label{tab:verdict}
\begin{tabular}{p{0.20\textwidth} p{0.13\textwidth} p{0.10\textwidth}
p{0.13\textwidth} p{0.10\textwidth} p{0.14\textwidth}}
\toprule
Estimator & Mean $|$error$|$, both experiments & Interval covers both &
Interval width (positive) & Needs geo panel & Discovers the intervention \\
\midrule
Pre-post & 6.5\% & - & - & no & no \\
Static regression & 22.9\% & no & 1,740,715 & no & no \\
Free level (BSTS) & 5.7\% & yes & 12,819,158 & no & no \\
Proposed & 13.0\% & yes & 5,701,130 & no & yes \\
Difference-in-differences & 22.3\% & - & - & yes & no \\
Synthetic control & 29.0\% & - & - & yes & no \\
\bottomrule
\end{tabular}
\par\smallskip
{\footnotesize\itshape Pre-post's low mean error carries no interval and does
not survive the simulations of Section~\ref{sec:sim}; see the text.}
\end{table}

\begin{figure}[t]
\centering
\includegraphics[width=\textwidth]{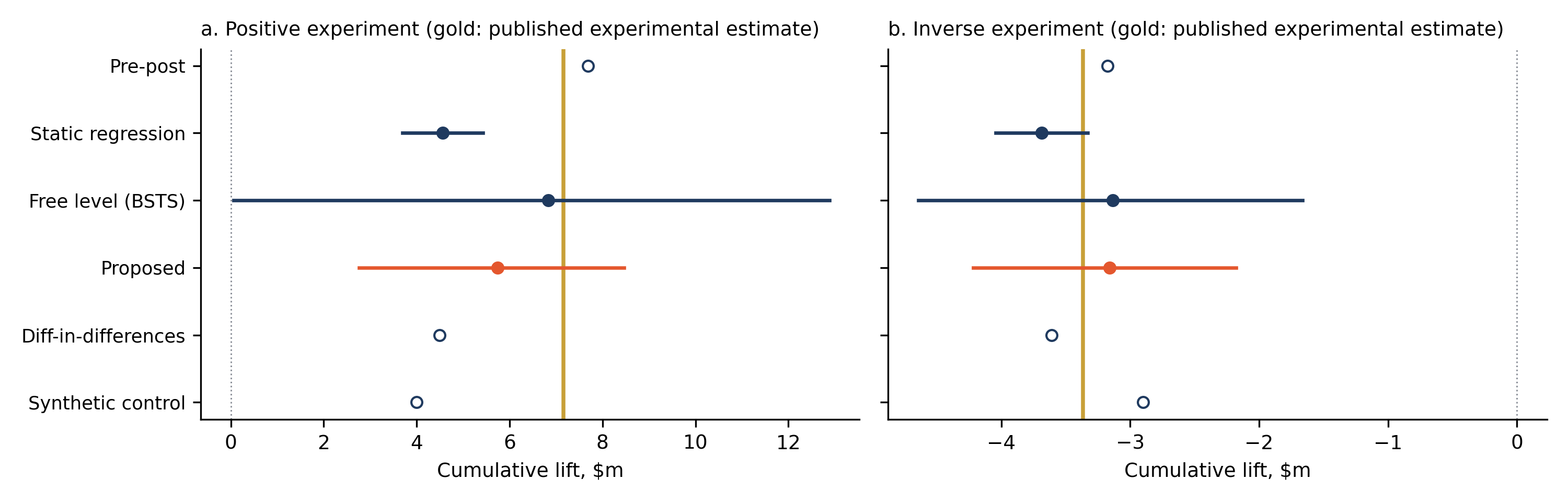}
\caption{The verdict at a glance: all six estimators against the two
published experimental estimates (gold), 90\% intervals where the method
produces one, proposed estimator in coral.}\label{fig:forest}
\end{figure}

Table~\ref{tab:verdict} collects the criteria and Figure~\ref{fig:forest}
displays the estimates; the question deserves a direct answer. On point error
alone the ranking is mixed: the free-level variant and pre-post average under
7 percent across the two experiments, the proposed estimator averages 13
percent, and the panel methods average 22 to 29 percent. But point error
alone is the wrong criterion, for stated reasons: pre-post offers no interval
and collapses in Section~\ref{sec:sim}'s simulations, and the free-level
variant's good points arrive inside intervals so wide that they assert almost
nothing, which is what a high-variance estimator looks like when a favorable
draw lands near the truth. On the criteria a budget decision actually
consumes, the answer is yes, the proposed procedure wins. Two estimators
produce intervals that cover both experimental estimates and exclude zero,
the proposed and the free-level variant; the proposed intervals are 2.2
and 1.5 times narrower, and narrowness is the entire value of an interval
that is already correct. The proposed procedure is also the only one of any
kind that dates the intervention it evaluates from the spending data alone
and that, in Section~\ref{sec:sim}'s simulations, discovers interventions
nobody told it about. The verdict carries two qualifications. An analyst who
needs only a point estimate on the positive experiment would have done better
with the free-level variant on this draw; we regard that trade as a bad one,
because decisions consume intervals, not points. And nothing here beats a
randomized experiment evaluated by its own design: the benchmark itself is
such an analysis, and where a firm can run one, it should. The claim this
article defends is narrower and, we think, more useful: when the experiment
is the one the firm already ran by accident, the proposed procedure recovers
the experimental answer, with calibrated uncertainty, from two series and a
spending schedule.

\section{Discussion and Conclusion}\label{sec:conc}
The procedure widens what can be analyzed as an experiment. Retrospectively,
a manager can price interventions that already happened, on channels for
which no lift study exists or ever will, at zero media cost and no risk,
because the intervention is already over. Prospectively, any planned budget
change becomes a designed quasi-experiment, with C1 through C5 acting as a
pre-registration discipline and the power bound \eqref{eq:rhomin} pricing the
design before commitment; this addresses directly the documented waste of
spending on underpowered experiments \citep{lewisrao15}. Qualifying estimates
feed downstream uses, including the experimental-prior calibration interfaces
of Bayesian marketing mix models, a practice with an open debate of its own
\citep{venkatraman25}, which we deliberately leave aside here.

The chief limitation is the source of ground truth. The Meta benchmarks,
though published as reference analyses and verifiable from their files, are
simulated, and the field lacks a public corpus pairing real advertiser series
with the lift studies run on them; the simulations show, moreover, that when
the control set is weak all regression-based estimators share the same bias,
so qualification can refuse a test but cannot manufacture information the
data do not contain. The positive experiment's 20 percent underestimate is
the price of the identification constraint on a series whose September
dynamics the untreated markets only partly carry, and we report it rather
than tune it away. Building a public corpus of real paired experiments,
extending detection to multivariate and gradual interventions, and mapping
the operating characteristics over a wider battery of designs are the
priorities for future work.

The conclusion we press is methodological: where experimentation is rationed,
the interventions firms make anyway are an estimable source of the same kind
of evidence, provided discovery, counterfactual construction, and
qualification are treated as one statistical procedure rather than three
afterthoughts.

\section*{Disclosure Statement}
\ifidentified
The author is the founder of MarSci Limited, a marketing analytics
consultancy. No other competing interests are declared.
\else
One of the authors is the founder of a marketing analytics consultancy. No other
competing interests are declared.
\fi

\section*{Funding}
No funding was obtained for this work.

\section*{Data Availability Statement}
The Meta benchmark datasets are public (GeoLift repository). The simulated
markets are reproducible from open-source code and reported seeds. Complete
replication code is provided as supplementary material.

\begin{singlespace}

\end{singlespace}

\clearpage
\appendix
\section{Proof Sketch of Proposition 1}\label{app:proof}
Write the measurement equation over the pre-window in stacked form as $\log y
= Z\beta + e$, with composite error $e_t = H\alpha_t + \varepsilon_t$ and
covariance $\Omega$ fixed by the constraint \eqref{eq:constraint}. Given
$(\sigma_\eta, \sigma_\zeta, \sigma_\varepsilon)$, the maximizer of
\eqref{eq:lik} in $\beta$ is generalized least squares, $\hat\beta = A \log
y$ with $A = (Z'\Omega^{-1}Z)^{-1} Z'\Omega^{-1}$ and $A Z = I$; with
$\sigma_\varepsilon$ estimated jointly the argument holds conditionally on
the estimated scale. Under the true model $\log y = Z\beta + x_k \theta + e$,
so conditional on the regressors $\mathrm{E}[\hat\beta] = \beta + A x_k
\theta = \beta + \Gamma\theta$, and post-window predictions inherit the bias
$z_t'\Gamma\theta$. When the constants in \eqref{eq:constraint} are small,
$\Omega$ is dominated by the measurement variance and $\Gamma$ approaches the
ordinary least squares coefficient vector of $x_k$ on $z$. If $x_k = 0$
throughout the pre-window, then $\Gamma = 0$ and the contamination vanishes,
which is the content of Corollary 1; the simulation magnitudes are those
reported with the corollary in Section~\ref{sec:cf}.

\section{Tuning Constants}\label{app:constants}
\begin{table}[h]
\centering\small
\caption{All tuning constants of the procedure.}\label{tab:constants}
\begin{tabular}{p{0.14\textwidth} p{0.38\textwidth} p{0.38\textwidth}}
\toprule
Constant & Value & Role \\
\midrule
$h$ & $1/13$ per week & break hazard; one expected break per planning
quarter \\
$r^{*}, \lambda$ & 4, 0.6 & short-run mass declaration rule \\
$P$ & 52 weeks; all available pre-treatment days on daily data & pre-window
length \\
$W$ & 8 weeks; the documented intervention length on daily data &
post-window length \\
$c_\eta, c_\zeta$ & 0.01, 0.001 & state-variance constraints
(identification) \\
$M$ & 2,000 & simulated counterfactual paths \\
C1 & $R^2 \ge 0.7$; KS $p \ge 0.01$ & counterfactual adequacy \\
C2 to C4 & 20\%; $3\times$; 0.95 & materiality; dominance; directional
significance \\
\bottomrule
\end{tabular}
\end{table}

\end{document}